\documentclass[revtex4]{emulateapj}

\usepackage{natbib}
\usepackage{color}

\usepackage{ulem}

\usepackage{bm}
\usepackage{lscape}

\newcommand{\eg}{e.g., }
\newcommand{\ie}{i.e., }
\newcommand{\Msun}{M_{\odot}}
\newcommand{\kms}{km~s$^{-1}$}

\def\gsim{\mathrel{\rlap{\lower 4pt \hbox{\hskip 1pt $\sim$}}\raise 1pt
\hbox {$>$}}}
\def\lsim{\mathrel{\rlap{\lower 4pt \hbox{\hskip 1pt $\sim$}}\raise 1pt
\hbox {$<$}}}

\def\ion#1#2{{\rm #1}~{\sc #2}}

\shorttitle{Discovery of Dramatic Optical Variability in SDSS J1100+4421}
\shortauthors{Tanaka et al.}

\begin{document}

\title{Discovery of Dramatic Optical Variability in SDSS J1100+4421: \\
A Peculiar Radio-Loud Narrow-Line Seyfert 1 Galaxy?
}
\author{
Masaomi Tanaka\altaffilmark{1}, 
Tomoki Morokuma\altaffilmark{2},
Ryosuke Itoh\altaffilmark{3},
Hiroshi Akitaya\altaffilmark{4},
Nozomu Tominaga\altaffilmark{5,6},
Yoshihiko Saito\altaffilmark{7},
{\L}ukasz Stawarz\altaffilmark{8,9},
Yasuyuki T. Tanaka\altaffilmark{4},
Poshak Gandhi\altaffilmark{10},
Gamal Ali\altaffilmark{11},
Tsutomu Aoki\altaffilmark{12},
Carlos Contreras\altaffilmark{13},
Mamoru Doi\altaffilmark{2},
Ahmad Essam\altaffilmark{11},
Gamal Hamed\altaffilmark{11},
Eric Y. Hsiao\altaffilmark{13},
Ikuru Iwata\altaffilmark{14},
Koji S. Kawabata\altaffilmark{4},
Nobuyuki Kawai\altaffilmark{7},
Yuki Kikuchi\altaffilmark{2},
Naoto Kobayashi\altaffilmark{2},
Daisuke Kuroda\altaffilmark{15},
Hiroyuki Maehara\altaffilmark{11}, 
Emiko Matsumoto\altaffilmark{5}, 
Paolo A. Mazzali\altaffilmark{16,17,18},
Takeo Minezaki\altaffilmark{2},
Hiroyuki Mito\altaffilmark{11}, 
Takashi Miyata\altaffilmark{2},
Satoshi Miyazaki\altaffilmark{1},
Kensho Mori\altaffilmark{3},
Yuki Moritani\altaffilmark{4},
Kana Morokuma-Matsui\altaffilmark{19},
Nidia Morrell\altaffilmark{13}, 
Tohru Nagao\altaffilmark{20}, 
Yoshikazu Nakada\altaffilmark{2},
Fumiaki Nakata\altaffilmark{14},
Chinami Noma\altaffilmark{21},
Ken Ohsuga\altaffilmark{1},
Norio Okada\altaffilmark{1},
Mark M. Phillips\altaffilmark{13}, 
Elena Pian\altaffilmark{22,23}, 
Michael W. Richmond\altaffilmark{24},
Devendra Sahu\altaffilmark{25},
Shigeyuki Sako\altaffilmark{2},
Yuki Sarugaku\altaffilmark{8},
Takumi Shibata\altaffilmark{5},
Takao Soyano\altaffilmark{11}, 
Maximilian D. Stritzinger\altaffilmark{26},
Yutaro Tachibana\altaffilmark{7}, 
Francesco Taddia\altaffilmark{27}, 
Katsutoshi Takaki\altaffilmark{3},
Ali Takey\altaffilmark{11},
Ken'ichi Tarusawa\altaffilmark{12}, 
Takahiro Ui\altaffilmark{3},
Nobuharu Ukita\altaffilmark{15},
Yuji Urata\altaffilmark{28},
Emma S. Walker\altaffilmark{29},
Taketoshi Yoshii\altaffilmark{7}
}

\altaffiltext{1}{National Astronomical Observatory of Japan, Mitaka, Tokyo 181-8588, Japan; masaomi.tanaka@nao.ac.jp}
\altaffiltext{2}{Institute of Astronomy, School of Science, University of Tokyo, Mitaka, Tokyo 181-0015, Japan}
\altaffiltext{3}{Department of Physical Sciences, Hiroshima University, Higashi-Hiroshima, Hiroshima 739-8526, Japan}
\altaffiltext{4}{Hiroshima Astrophysical Science Center, Hiroshima University, Higashi-Hiroshima, Hiroshima 739-8526, Japan}
\altaffiltext{5}{Department of Physics, Faculty of Science and Engineering, Konan University, Kobe, Hyogo 658-8501, Japan}
\altaffiltext{6}{Kavli Institute for the Physics and Mathematics of the Universe (WPI), The University of Tokyo, Kashiwa, Chiba 277-8583, Japan}
\altaffiltext{7}{Department of Physics, Tokyo Institute of Technology, Meguro-ku, Tokyo 152-8551, Japan}
\altaffiltext{8}{Institute of Space and Astronautical Science, JAXA, Sagamihara, Kanagawa 252-5210, Japan}
\altaffiltext{9}{Astronomical Observatory, Jagiellonian University, ul. Orla 171, 30-244 Krakow, Poland}
\altaffiltext{10}{Department of Physics, Durham University, Durham DH1-3LE, UK}
\altaffiltext{11}{National Research Institute of Astronomy and Geophysics, Helwan, Cairo, Egypt}
\altaffiltext{12}{Kiso Observatory, Institute of Astronomy, School of Science, The University of Tokyo, Kiso, Nagano 397-0101, Japan}
\altaffiltext{13}{Carnegie Observatories, Las Campanas Observatory, Colina El Pino, Casilla 601, Chile}
\altaffiltext{14}{Subaru Telescope, National Astronomical Observatory of Japan, Hilo, HI 96720, USA}
\altaffiltext{15}{Okayama Astrophysical Observatory, National Astronomical Observatory of Japan, Asakuchi, Okayama 719-0232, Japan}
\altaffiltext{16}{Astrophysics Research Institute, Liverpool John Moores University, IC2, Liverpool Science Park, 146 Brownlow Hill, Liverpool L3 5RF, UK}
\altaffiltext{17}{Istituto Nazionale di Astrofisica-OAPd, vicolo dell'Osservatorio 5, I-35122 Padova, Italy}
\altaffiltext{18}{Max-Planck-Institut f\"{u}r Astrophysik, Karl-Schwarzschild-Str. 1, D-85748 Garching, Germany}
\altaffiltext{19}{Nobeyama Radio Observatory, Nobeyama, Minamimaki, Minamisaku, Nagano 384-1305, Japan}
\altaffiltext{20}{Research Center for Space and Cosmic Evolution, Ehime University, Bunkyo-cho, Matsuyama 790-8577, Japan}
\altaffiltext{21}{Astronomical Institute, Tohoku University, Aramaki, Aoba-ku, Sendai 980-8578, Japan}
\altaffiltext{22}{Scuola Normale Superiore di Pisa, Piazza dei Cavalieri 7, I-56126 Pisa, Italy}
\altaffiltext{23}{INAF-Istituto di Astrofisica Spaziale e Fisica Cosmica, Via P. Gobetti 101, I-40129 Bologna, Italy}
\altaffiltext{24}{Department of Physics, Rochester Institute of Technology, 85 Lomb Memorial Drive, Rochester, NY 14623-5603, USA}
\altaffiltext{25}{Indian Institute of Astrophysics, Koramangala, Bangalore 560 034, India}
\altaffiltext{26}{Department of Physics and Astronomy, Aarhus University, Ny Munkegade, DK-8000 Aarhus C, Denmark}
\altaffiltext{27}{The Oskar Klein Centre, Department of Astronomy, Stockholm University, AlbaNova, SE-10691 Stockholm, Sweden}
\altaffiltext{28}{Institute of Astronomy, National Central University, Chung-Li 32054, Taiwan}
\altaffiltext{29}{Department of Physics, Yale University, New Haven, CT 06520-8120, USA}

\begin{abstract}
We present our discovery of dramatic variability in
SDSS J1100+4421 by the high-cadence transient survey Kiso Supernova Survey 
(KISS). 
The source brightened in the optical by at least a factor of
three within about half a day.
Spectroscopic observations suggest that
this object is likely a narrow-line Seyfert 1 galaxy (NLS1) at
$z=0.840$, however with unusually strong narrow emission lines.
The estimated black hole mass of $\sim 10^7 \Msun$ implies 
bolometric nuclear luminosity close to the Eddington limit. 
SDSS J1100+4421 is also extremely radio-loud, 
with a radio loudness parameter of 
$R \simeq 4 \times 10^2 - 3 \times 10^3$,
which implies the presence of relativistic jets. 
Rapid and large-amplitude optical
variability of the target, reminiscent of that found in a few 
radio- and $\gamma$-ray loud NLS1s, 
is therefore produced most likely in a blazar-like core. 
The 1.4 GHz radio image of the
source shows an extended structure with a linear size of about 100 kpc. 
If SDSS J1100+4421 is a genuine NLS1, as suggested here,
this radio structure would then be the largest ever discovered in this
type of active galaxies.
\end{abstract}


\begin{figure*}
\begin{center}
\begin{tabular}{lc}
\begin{minipage}{0.2\textwidth}
\includegraphics[height=0.5\textheight]{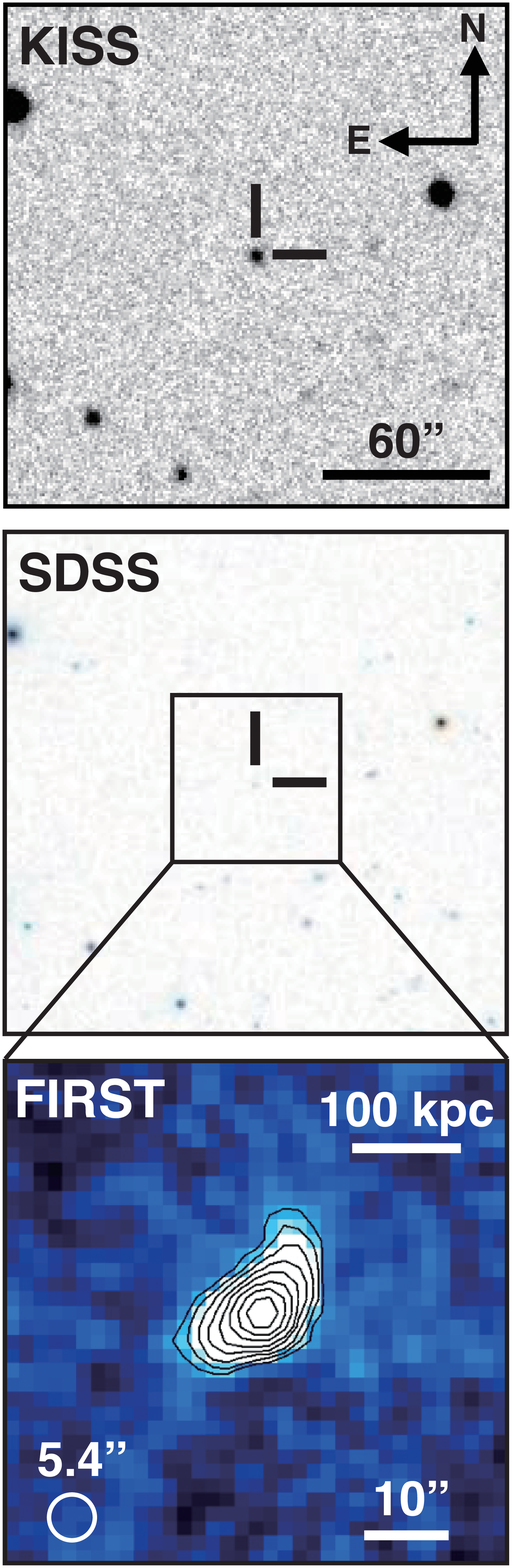}
\end{minipage}
&
\begin{minipage}{0.65\textwidth}
\centering
\includegraphics[]{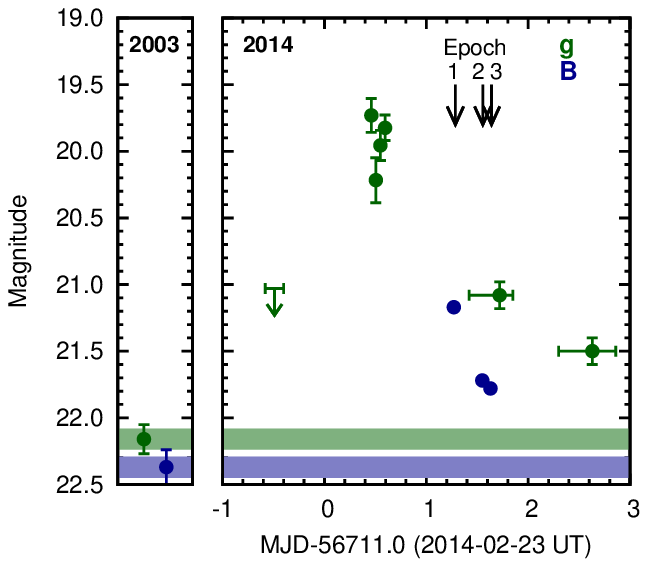} \\
\includegraphics[]{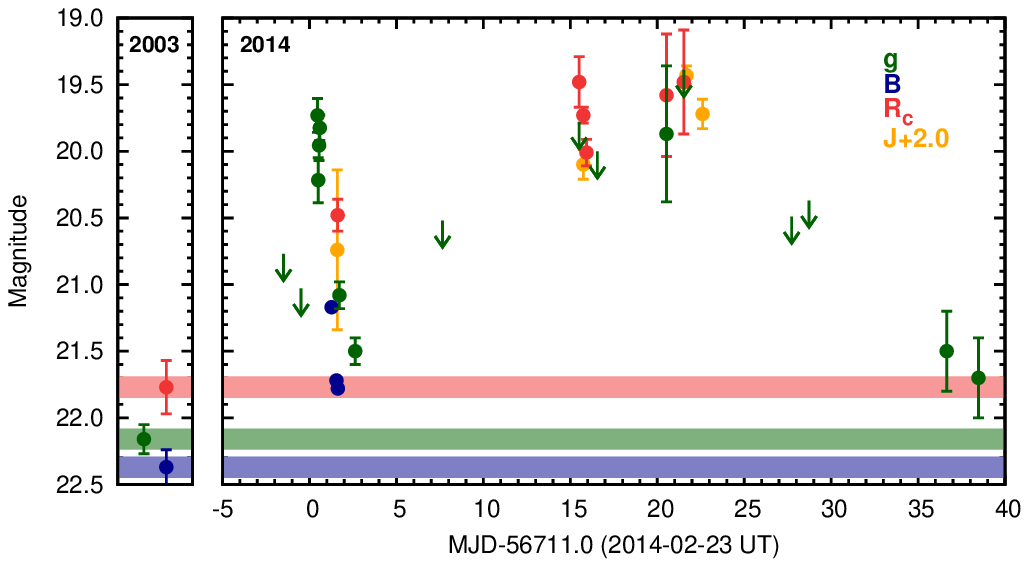}
\end{minipage}
\end{tabular}
\caption{
{\it Left}: Discovery image of SDSS J1100+4421 taken 
on 2014 Feb 23 UT with KWFC (upper),
the SDSS image (middle),
and the FIRST 1.4 GHz radio image (lower, the beam size is 5.4'' as shown 
with the circle, \citealt{white97}).
The contour levels in the FIRST image 
start from 3$\sigma$ (1 $\sigma =$ 0.20 mJy beam$^{-1}$),
separated by a factor of $\sqrt{2}$.
{\it Right}: Light curves of SDSS J1100+4421 
around the discovery epoch (upper) and until 40 days after the flare (lower).
The vertical arrows in black show the epochs of spectroscopic data.
Our photometric data are compared with SDSS $g$-band magnitude
and $B$- and $R_c$-band magnitudes estimated from SDSS magnitudes
(left panel and shaded region).
Magnitudes are given in the AB system for the SDSS $g$ band
and Vega units for the other bands.
}
\label{fig:discovery}
\end{center}
\end{figure*}

\section{Introduction}
\label{sec:intro}

It is widely accepted that active galactic nuclei (AGNs) 
are powered by supermassive black holes (BHs) accreting at high rates.
Radio-loud AGNs are those which possess powerful relativistic jets.
The radio loudness parameter $R$, \ie the ratio of a radio flux to 
a nuclear optical flux of a source, is often used as a proxy for 
the jet production efficiency. Several plausible relations between this
parameter and other fundamental characteristics of a central engine
(such as BH mass, accretion rate, and possibly BH spin) have been
discussed in the literature (e.g., \citealt{sikora07}, and references therein).
In general, AGNs with higher BH masses ($> 10^8 \Msun$) 
and lower accretion rates tend to be more radio loud.

In this context, radio properties of NLS1s 
are of a particular interest, since AGNs of this peculiar type 
\citep{osterbrock85,pogge00}
are believed to have relatively small BH masses ($10^6 - 10^8 \Msun$) 
and very high accretion rates \citep[\eg][]{mathur00}.
By these properties, it had been inferred that NLS1s are a radio-quiet 
class of AGNs, and that young BHs in NLS1s that undergo rapid growth via 
high-rate accretion do not produce relativistic jets.

Statistical studies 
(\eg \citealt{komossa06}; \citealt{whalen06}; \citealt{zhou06} hereafter Z06)
show, in fact, that the fraction of radio-loud NLS1s is small,
\ie only 7\% of NLS1s have $R>10$ while about 20 \% of 
broad-line Seyfert 1 galaxies have $R>10$ \citep{komossa06}.
However, they also find that a small fraction ($\sim 2.5 \%$) of NLS1s 
is very radio-loud ($R>100$).
Recently more and more radio-loud NLS1s are being discovered 
\citep[\eg][]{yuan08,caccianiga14}.
Interestingly, high-energy $\gamma$-rays (100 MeV - 100 GeV) have been 
detected from some radio-loud NLS1s with {\it Fermi} Large Area Telescope
\citep[LAT;][]{abdo09PMN0948,abdo09,dammando12}.
The $\gamma$-ray detection implies the presence of relativistic jets
in these objects, in direct analogy to blazars.
To fully understand the cosmological evolution of supermassive BHs, 
it is therefore important to clarify the jet production efficiency and
the jet duty cycle in such young systems in formation, or in other
words to investigate in detail the radio-loud population of NLS1s.

In this Letter, we report our serendipitous discovery of 
dramatic optical variability in a peculiar
radio-loud NLS1 candidate SDSS J110006.07+442144.3 (SDSS J1100+4421)
by the Kiso Supernova Survey (KISS, \citealt{morokuma14}).
Throughout this paper, we assume cosmological parameters 
$\Omega_{\Lambda} = 0.7, \Omega_{M} = 0.3$, and $h=0.7$.

\section{Observations}
\label{sec:observations}

\subsection{Discovery}
\label{sec:discovery}

SDSS J1100+4421 was detected as a transient object
by the high-cadence optical transient survey KISS \citep{morokuma14}.
KISS uses the 1.05m Kiso Schmidt telescope equipped with  
Kiso Wide Field Camera \citep[KWFC,][]{sako12}, 
which has a 2.2 $\times$ 2.2 deg field of view.
In order to detect short-timescale transients, 
KISS adopts a 1 hr cadence; 
\ie the same fields are repeatedly observed every 1 hr.
The survey is performed with the optical $g$-band filter 
to detect shock breakout of supernovae \citep{tominaga11,morokuma14}.

Variability of SDSS J1100+4421 was first recognized on 2014 Feb 23.46 UT
with $g = 19.73 \pm 0.13$ mag 
(Figure \ref{fig:discovery} and Table \ref{tab:data})
and registered as a supernova candidate ``KISS14k''.
No source was present at this position in the stacked images taken 
1 day before with a 5-$\sigma$ limiting magnitude of 21.03 mag.
Therefore this transient brightened at least by a factor of 3
within 1 day (1.3 mag day$^{-1}$).
The SDSS images taken in 2003 show a faint object 
classified as a galaxy at the same position
with $g = 22.16 \pm 0.11$ mag without spectroscopic data.
Compared with the 2003 data,
the 2014 flux increase is a factor of about 10.
Hereafter we call this event a ``flare''.
This object is also recorded in USNO-B1.0 \citep{monet03}
and in the Guide Star Catalog II \citep[GSC-II,][]{lasker08}
with variable magnitudes (Table \ref{tab:data}).

After the flare, 
the brightness of SDSS J1100+4421 quickly declined in subsequent days.
The stacked images show the fading object with 
about $g = $21.1 and 21.5 mag on 2014 Feb 24 and 25, respectively.
The decline rate during the first day after the flare is
$0.9-1.3$ mag day$^{-1}$.

\begin{figure}
\begin{center}
\includegraphics[scale=1.2]{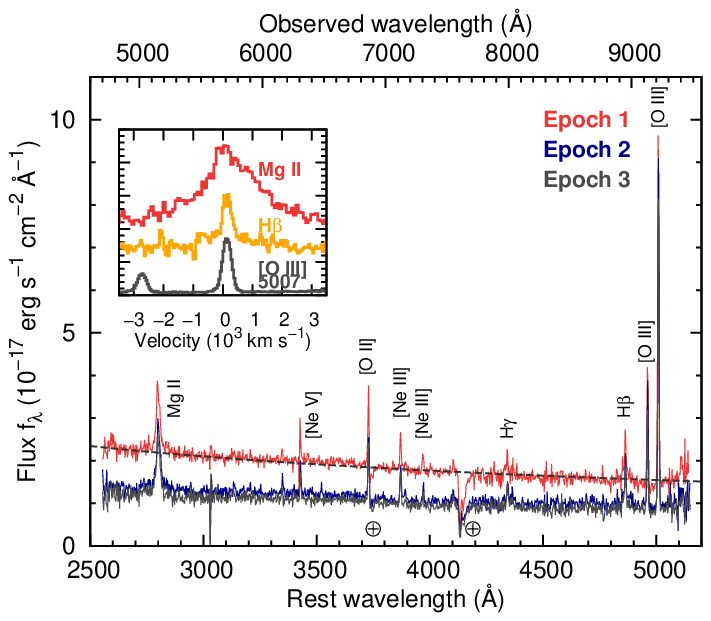} 
\includegraphics[scale=1.2]{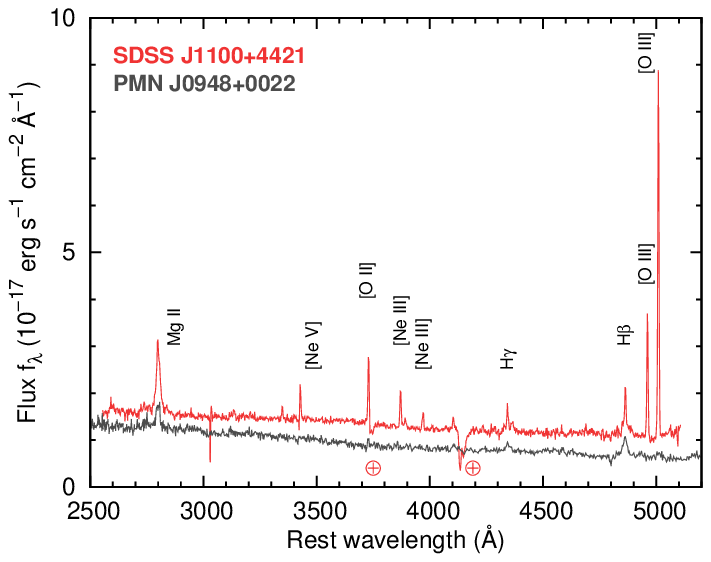} 
\caption{
{\it Upper}: 
Optical spectra of SDSS J1100+4421 taken with Subaru/FOCAS 
on 2014 Feb 24 UT. 
Epochs 1, 2, and 3 correspond to 
MJD = 56712.29, 56712.55, and 56712.64, respectively.
All the spectra are shown with 3 pixel binning (4.2 \AA\ per binned pixel).
The dashed line shows a power law fit with $f_{\lambda}\propto \lambda^{-0.6}$.
Profiles of strong emission lines in the 
stacked spectrum are shown in the inset 
(in the original pixel scale).
{\it Lower:} Stacked spectrum of SDSS J1100+4421 
(3 pixel binning)
compared with the SDSS spectrum of PMN J0948+0022 (flux scaled).
}
\label{fig:spec}
\end{center}
\end{figure}

\subsection{Follow-up Observations}
\label{sec:followup}

Immediately after the discovery (19.4 hr after the first detection),
we started follow-up imaging and spectroscopic observations
with the 8.2m Subaru telescope equipped with 
the Faint Object Camera And Spectrograph \citep[FOCAS,][]{kashikawa02}.
We took images and spectra of SDSS J1100+4421 
three times during the same night.
Typical seeing during the observations was 0$\farcs 5$-0$\farcs 6$.
Imaging data clearly show intranight variability.

Spectroscopic observations were performed using 
an offset 1.0'' slit with the 300B grism and SY47 order-sorting filter
\footnote{This configuration gives
clean 1st-order spectra in 4700-9000\AA. 
Contamination of the 2nd-order spectrum exists at $>$9000 \AA\ 
and it is estimated to be about 5\% at 9200 \AA\ 
for the spectra of SDSS J1100+4421.}.
The spectral resolution is $\lambda/\Delta\lambda = 400$ (FWHM),
which corresponds to a velocity resolution of 750 \kms. 
The obtained flux is consistent at a level of about 10\% 
with the FOCAS photometry.

Figure \ref{fig:spec} shows the optical spectra of SDSS J1100+4421.
Emission lines of \ion{Mg}{ii}, [\ion{Ne}{v}], [\ion{O}{ii}], 
[\ion{Ne}{iii}], H $\gamma$, H $\beta$, and [\ion{O}{iii]} 
are clearly identified, confirming the AGN nature of this object.
All the emission lines consistently indicate a redshift of $z = 0.840$.
All three spectra show a broad \ion{Mg}{ii} line
with a FWHM of $2,070\pm 100$ \kms, which does not change with time.
Although the H$\beta$ line is dominated by the narrow component,
the broad component is visible 
in the stacked spectrum (bottom panel of Figure \ref{fig:spec}).
The FWHM of the broad H$\beta$ is $1900 \pm 300$ \kms.
The widths of all the other narrow emission lines are not resolved 
with our spectra.

The continuum flux in the three spectra 
shows significant intranight variability.
We find that the continuum of these spectra is well fitted 
by a power-law with a slope that is constant in time.
The power-law index is
$\alpha_{\lambda} =-0.60 \pm 0.01$ in all three epochs,
where $f_{\lambda}\propto \lambda^{\alpha_{\lambda}}$,
($\alpha_{\nu} = -1.4$ for $f_{\nu} \propto \nu^{\alpha_{\nu}}$).
In contrast to the continuum flux,
the emission line fluxes in the three spectra
are consistent within the uncertainty,
\eg the variation in the \ion{Mg}{ii} line flux is $\lsim 10 \%$.
Emission line fluxes measured in the stacked spectrum 
are summarized in Table \ref{tab:data}.

Optical and near infrared imaging follow-up observations 
were also performed 
with Horishima Optical and Near-InfraRed camera (HONIR, \citealt{akitaya14}) 
of the 1.5m Kanata telescope,
the Newtonian camera of the Kottamia Observatory 1.88m telescope,
the Kyoto Okayama Optical Low-dispersion Spectrograph (KOOLS)
of the Okayama Astrophysical Observatory (OAO) 
1.88m telescope \citep{yoshida05},
and the 0.50m MITSuME telescope \citep{yatsu07}.
Even after the flare, this object shows some variability 
(Figure \ref{fig:discovery}),
although the data are not densely sampled.
The $r$-$i$ color during the flare (synthesized from the FOCAS spectra) 
is 0.25 mag, which is consistent with the SDSS $r$-$i$ color in 2003 
($0.26 \pm 0.21$ mag).

In order to measure the X-ray flux of this object,
Target of Opportunity (ToO) observations 
were performed with the Swift XRT \citep{burrows05XRT} on 2014 Mar 16-17 UT.
An X-ray source was detected at a position consistent with SDSS J1100+4421.
Assuming a power-law photon index of $-2.0$
(as for PMN J0948+0222, \citealt{abdo09PMN0948}),
the average 0.3--8 keV flux reads as  
$(1.62 \pm 0.38) \times 10^{-13} \ {\rm erg \ s^{-1} \ cm^{-2}}$.

\section{Discussion}
\label{sec:discussion}

\subsection{Nature of SDSS J1100+4421}
\label{sec:nature}

Our optical spectroscopy clearly indicates an AGN nature for
SDSS J1100+4421. The relatively narrow widths of the broad \ion{Mg}{ii} and
H$\beta$ lines, FWHM $\lsim 2000$\ km\ s$^{-1}$, suggest in addition a
NLS1 classification \citep[\eg][]{osterbrock85,pogge00,komossa06}
\footnote{
Note that the dramatic optical variability observed, as well as the
clearly detected broad components of the \ion{Mg}{ii} and H$\beta$ lines,
both invalidate a Seyfert 2 classification of SDSS J1100+4421. 
Also, the hardness ratio of the X-ray spectrum of the source, 
$(H - S)/(H + S) = -0.14 \pm 0.13$ (where $H$ and $S$ are the photon counts 
at $< 2$\,keV and $> 2$\,keV, respectively), does not indicate any heavy
nuclear obscuration which could be expected for a type 2 AGN \citep{wangjx04}.}.
However, SDSS J1100+4421 does not fit the other
two characteristics of NLS1s, namely (i) the presence of Fe II bumps
at 4400-4700\,$\AA$ and 5150-5400\,$\AA$, and 
(ii) [\ion{O}{iii}]/H$\beta$ flux ratio $< 3$.

The absence of a \ion{Fe}{ii} bump 
may be naturally explained by a contribution from the additional
prominent continuum component during the flare.
This component can be interpreted as emission 
from a jet (Section \ref{sec:flare}).
In fact, this object is found to be extremely radio loud
(Section \ref{sec:radio}), and radio-loud AGNs tend to show weaker 
\ion{Fe}{ii} bumps \citep{boroson92,sulentic03}.

Figure \ref{fig:spec} (bottom panel) shows that the spectrum of 
SDSS J1100+4421 is similar to that of NLS1 PMN J0948+0022
except for the narrow emission lines such as [\ion{O}{ii}] and [\ion{O}{iii}].
In fact, considering the properties of the broad emission lines and
continuum, the narrow [\ion{O}{iii}] line is unusually strong, 
with the corresponding [\ion{O}{iii}]/H$\beta$ flux ratio $\simeq 5-9$.
Assuming a conventional conversion
($L_{5100} \simeq 10^{2.5} L_{\rm [O\ III]}$ and 
$L_{\rm bol} = 8.1 L_{5100}$, \citealt{shen11,runnoe12}),
the [\ion{O}{iii}] line luminosity gives an estimate for the bolometric
AGN luminosity $L_{\rm bol} \simeq 8 \times 10^{45} \ { \rm erg\ s^{-1}}$. 
This is about 10 times higher than that
estimated using the broad \ion{Mg}{ii} line, or the SDSS g-band flux in 2003,
$L_{\rm bol} = 5.2 L_{\rm 3000} \simeq 6 \times 10^{44} \ {\rm erg \ s^{-1}}$.

Based on these findings and considerations, we conclude that
SDSS J1100+4421 is a candidate NLS1 with a particularly prominent
power-law component and unusually strong narrow emission lines.
Since the [\ion{O}{iii}] line strength is known to correlate
with the radio power \citep[\eg][]{labiano08}, 
the strong narrow components might be related to
the extreme radio loudness and thus relativistic jets (Section \ref{sec:radio}).

\subsection{Radio Properties}
\label{sec:radio}

SDSS\,J1100+4421 is a strong radio source, which is recorded in
various survey catalogs (Table \ref{tab:data}). 
For a direct comparison with NLS1s,
here we define a radio loudness parameter $R_{1.4}$ as in 
Z06, \citet{komossa06} and \citet{yuan08}, 
\ie as the ratio of $f_{\nu}$ (1.4\,GHz) to $f_{\nu}$ (4400 \AA) 
in the source rest frame
(assuming radio and optical power-law slopes of $\alpha_{\nu} = -0.5$
for simple flux conversions).

\begin{figure}
\begin{center}
\includegraphics[scale=1.0]{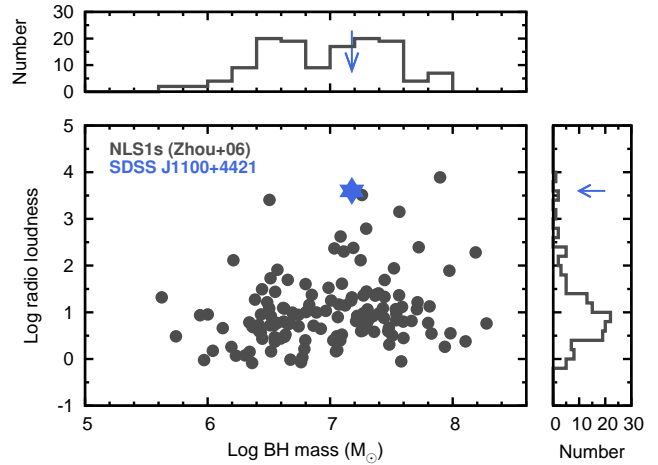} 
\caption{
Radio loudness $R_{1.4}$ and BH mass for NLS1 samples from Z06
and for SDSS J1100+4421.
The right and top panels show the histograms of radio loudness and BH mass,
respectively. 
The values for SDSS J1100+4421 are shown with arrows.}
\label{fig:BH_RL}
\end{center}
\end{figure}

Using the radio flux measured by FIRST
and the SDSS $g$-band magnitude corrected for Galactic extinction
($A_V = 0.035$ mag, \citealt{schlegel98}),
the radio loudness is $R_{1.4} \simeq 3 \times 10^3$.
This is among the highest radio loudness values in the NLS1 samples 
(Figure \ref{fig:BH_RL}).
It should be noted that the epochs of various radio observations of
SDSS J1100+4421 are not simultaneous with each other, or with those
in other wavelengths; hence the evaluated radio loudness parameter is
subject to an uncertainty generated by a source variability.
However, even with the optical flux at the peak of the flare, 
the radio loudness is still very high, $R_{1.4} \simeq 4 \times 10^2$.

Interestingly, the radio structure in the FIRST image is extended
compared with the beam size (Figure \ref{fig:discovery}).
The source has a core-dominated, two-sided structure
with the linear size of 13'' 
(measured from the center to the north-west edge with 3$\sigma$ flux level).
This size corresponds to about 100 kpc (projected).
If SDSS J1100+4421 is a NLS1, such a large-scale radio structure would
be the largest known for this type of AGN 
\citep[][and references therein]{doi12}.
Future high-resolution observations are needed to confirm the
extension, and to investigate in detail the radio morphology of
SDSS J1100+4421.

\subsection{Black Hole Mass and Eddington Ratio}
\label{sec:BH}

We estimate a mass of the BH in SDSS J1100+4421 with 
the following conventional method.
Assuming the broad-line region (BLR) is virialized,
the black hole mass can be written as
$M_{\rm BH} = R_{\rm BLR} v^2 /G$, 
where $R_{\rm BLR}$ is the size of BLR 
and $v = (\sqrt{3} \times \rm{FWHM}/2)$ for 
an isotropic distribution of BLR clouds.
The size of BLR ($R_{\rm BLR}$) is known to correlate with
the continuum and line luminosities \citep[\eg][]{kaspi05}.
We use the empirical relations of \citet[their Eqs A6 and A7]{mclure04},
which give consistent BH mass estimates within 0.33 dex
using H$\beta$ and \ion{Mg}{ii}.
To avoid possible contamination from jet emission \citep{yuan08},
the continuum luminosities in the equations
are replaced with the line luminosities
by adopting the relations of \citet[$L_{3000}$ and $L_{\rm Mg\ II}$]{shen11} 
and \citet[$L_{5100}$ and $L_{\rm H\beta}$]{greene05}, respectively.

By adopting the luminosity and FWHM of the 
H$\beta$ and \ion{Mg}{ii} lines,
we obtain $M_{\rm BH} \sim 1.0 \times 10^7 \Msun$
and $1.5 \times 10^7 \Msun$, respectively.
Given the higher S/N ratio, 
we adopt the estimate with the \ion{Mg}{ii} line 
in the following discussion.
Figure \ref{fig:BH_RL} shows
the radio loudness and BH mass compared with those of NLS1 samples by Z06.
The BH masses for this sample are estimated using 
the luminosity and FWHM of the H$\beta$ line.
The BH mass of SDSS J1100+4421 is 
within the distribution of BH masses of other NLS1s.

The bolometric luminosity of SDSS J1100+4421 is
$L_{\rm bol} \sim 6 \times 10^{44} \ {\rm erg \ s^{-1}}$
(Section \ref{sec:nature}), which is about 30 \% of the Eddington luminosity
($L_{\rm Edd} \sim 2 \times 10^{45} \ {\rm erg \ s^{-1}}$).
Since radiative efficiency of the accretion disk at high 
(close to Eddington) accretion rates is of the order of 10\%, 
the accretion luminosity of SDSS J1100+4421 is approximately 
$L_{\rm acc} \sim 6 \times 10^{45} \ {\rm erg \ s^{-1}}$,
corresponding to the mass accretion rate of 
$\dot{M}_{\rm acc} = L_{\rm acc} / c^2 \simeq 0.1\ \Msun \ {\rm yr^{-1}}$.
The jet power estimated from the radio luminosity 
($L_{1.4} = 7 \times 10^{43}\ {\rm erg\ s^{-1}}$) is 
$L_{\rm jet} \sim 1 \times 10^{45}\ {\rm erg\ s^{-1}}$ \citep{cavagnolo10},
which confirms the high jet production efficiency.

\begin{figure}
\begin{center}
\includegraphics[scale=1.2]{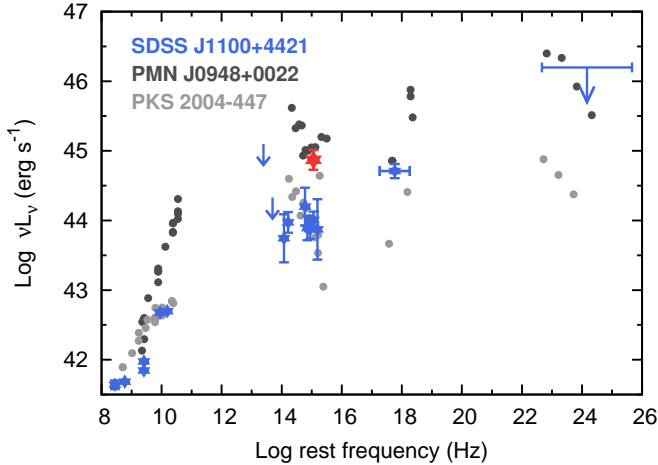} 
\caption{
SED of SDSS J1100+4421 (blue) compared with 
that of $\gamma$-ray loud NLS1s 
PMN J0948+0022 \citep{abdo09PMN0948}
and PKS 2004-447 \citep{gallo06,abdo09}.
The optical flux of SDSS J1100+4421 during the flare is shown in red.
}
\label{fig:sed}
\end{center}
\end{figure}

\subsection{Origin of the Flare}
\label{sec:flare}

The long-term light curve rejects the flare in SDSS J1100+4421 
as a transient phenomena, such as supernova or tidal disruption of a star.
One might suppose that the flare arises 
from the variability of an accretion disk, but that is also unlikely.
Assuming a standard (optically thick, geometrically thin) disk
structure around a BH with $M_{\rm BH} \sim 1.5 \times 10^7 \Msun$, 
the disk radius at which the bulk of the
optical/ultraviolet emission is produced can be estimated as 
$R \sim 10^{15}$\,cm\,$\sim 200 \,R_S$, where $R_S$ is the Schwarzschild radius.
This emitting radius is not changed significantly 
even for a super-critical, slim disk as long as $R > 10 R_{\rm S}$
\citep[\eg][]{mineshige00}.
The dynamical timescale at such a distance is a few tens of days,
which is much longer than the observed timescale.

The remaining possibility for the origin of the flare 
is synchrotron emission from a relativistic jet.
The presence of a jet is implied by the extreme radio loudness 
of the source.
The steep slope of the optical spectra during the flare 
($\alpha_{\nu} = -1.4$)
is also consistent with synchrotron emission.
Moreover, the observed variability timescale points out to the jet
origin of the flare as well.

Optical variability on timescales shorter than a few days
has been observed in other NLS1s 
\citep[\eg][]{klimek04,liu10,paliya13,maune13,eggen13,itoh13}.
During the flare analyzed here, SDSS J1100+4421 brightens by a factor of 
at least 3 within about half a day in the rest frame.
To our knowledge, such an extreme, blazar-like variability of a NLS1
has been confirmed for only $\gamma$-ray loud NLS1s
\citep{liu10,maune13,eggen13,itoh13}.

Motivated by these similarities,
the SED of SDSS J1100+4421 is compared with 
$\gamma$-ray loud NLS1s PMN J0948+0022 and PKS 2004-447 
\citep{abdo09PMN0948,abdo09} in Figure \ref{fig:sed}.
SDSS J1100+4421 is not detected in 
the Two Year Fermi-LAT catalog \citep[2FGL,][]{nolan12}.
If the $\gamma$-ray photon index is assumed to be $-2.5$, 
in the analogy to PMN J0948+0222, the upper limit for the flux is
$\sim 5 \times 10^{-12} \  {\rm erg \ s^{-1} \ cm^{-2}}$.
There is a large variety in the $\gamma$-ray loudness of the
LAT-detected NLS1s; the upper limit for the $\gamma$-ray luminosity of
SDSS J1100+4421 derived here is very close to the observed
$\gamma$-ray luminosity of PMN\,J0948+0022, but it is higher than the
luminosity of PKS\,2004-447 by a factor of about 30.

\section{Summary}
\label{sec:summary}

We report on the discovery of a dramatic optical variability 
from an enigmatic AGN, SDSS J1100+4421.
This object seems like a NLS1
but is peculiar for its strong narrow emission lines
and the large ($\sim 100$ kpc) extent of the radio structure.
The observational facts presented here suggest that 
this object has a young BH growing with an accretion rate 
close to the Eddington limit,
and that the relativistic jets are efficiently produced
and have evolved to the large scale.
The dramatic optical variability,
as well as the unusually strong narrow emission lines,
are likely to be produced by the relativistic jets. 
Our discovery demonstrates that high-cadence surveys are potentially useful 
to search for such a rare class of AGNs
and to study the jet production/duty cycle in the growing BHs.

\acknowledgments
We thank the OAO staff for continuous support and 
the {\it Swift} team for approving our ToO request 
and performing the observations.
This research is based in part on data collected at Subaru Telescope, 
which is operated by the National Astronomical Observatory of Japan.
This research has made use of the NASA/IPAC Extragalactic Database (NED)
which is operated by the Jet Propulsion Laboratory, 
California Institute of Technology, 
under contract with the National Aeronautics and Space Administration.
The National Radio Astronomy Observatory is a facility of 
the National Science Foundation operated under cooperative agreement 
by Associated Universities, Inc.
This research has been supported by the Grant-in-Aid for Scientific Research 
(23740143, 23740157, 24740117, 25103515, 25800103)
of the Japan Society for the Promotion of Science (JSPS)
and of the Ministry of Education, Culture, Sports, Science and Technology 
(MEXT) and by INAF PRIN 2011 and PRIN MIUR 2010/2011.

\begin{deluxetable*}{cccccc} 
\tablewidth{0pt}
\tablecaption{Summary of New and Archival Data for SDSS J1100+4421}
\tablehead{ & & & & & }
\startdata
MJD$^{a}$   & UT$^{a}$  & Filter &  Magnitude$^{b}$ & Exp. time &  Telescope$^{c}$ \\
           &           &        &            &   (sec)   &   \\ \hline    
56325.66 & 2013 Feb 02  & $g$  &  $>$20.65$^{d}$    & $180 \times 5$ & 1 \\
56709.50 & 2014 Feb 21 & $g$ &  $>$20.77$^{d}$    & $180 \times 5$ & 1 \\
56710.51 & 2014 Feb 22 & $g$ &  $>$21.03$^{d}$    & $180 \times 5$ & 1 \\
56711.46 & 2014 Feb 23 & $g$ &  19.73 $\pm$ 0.13  & 180            & 1   \\
56711.51 & 2014 Feb 23 & $g$ &  20.22 $\pm$ 0.17  & 180            & 1   \\
56711.55 & 2014 Feb 23 & $g$ &  19.96 $\pm$ 0.11  & 180            & 1   \\
56711.60 & 2014 Feb 23 & $g$ &  19.82 $\pm$ 0.10  & 180            & 1   \\
56712.72 & 2014 Feb 24 & $g$ &  21.08 $\pm$ 0.10  & $180 \times 41$ & 1   \\
56713.63 & 2014 Feb 25 & $g$ &  21.50 $\pm$ 0.10  & $180 \times 44$ & 1   \\
56718.65 & 2014 Mar 02  & $g$ &  $>$20.52$^{d}$    & $180 \times 8$ & 1 \\
56726.51 & 2014 Mar 10 & $g$ &  $>$19.78$^{d}$    &  $60 \times 27$ & 2 \\
56727.56 & 2014 Mar 11 & $g$ &  $>$20.00$^{d}$    &  $180 \times 36$ & 1 \\
56731.53 & 2014 Mar 15 & $g$ &  19.87 $\pm$ 0.51 &  $60 \times 26$ & 2 \\
56732.53 & 2014 Mar 16 & $g$ &  $>$19.39$^{d}$    &  $60 \times 54$ & 2 \\
56738.73 & 2014 Mar 22 & $g$ &  $>$20.49$^{d}$    &  $180 \times 4$ & 1 \\
56739.73 & 2014 Mar 23 & $g$ &  $>$20.37$^{d}$    &  $180 \times 5$ & 1 \\ 
56747.65 & 2014 Mar 31 & $g$ &  21.50 $\pm$ 0.30  &  $180 \times 13$ & 3 \\ 
56749.47 & 2014 Apr 02  & $g$ &  21.70 $\pm$ 0.30  &  $180 \times 6 $ & 3 \\ \\

56712.63 & 2014 Feb 24 & $i$ &  20.72 $\pm$ 0.01  & 30             & 4 \\ \\

56712.27 & 2014 Feb 24 & $B$ &  21.17 $\pm$ 0.02  & 25             & 4 \\
56712.55 & 2014 Feb 24 & $B$ &  21.72 $\pm$ 0.02  & 15             & 4 \\
56712.63 & 2014 Feb 24 & $B$ &  21.78 $\pm$ 0.02  & 30             & 4 \\ \\

56726.93 & 2014 Mar 10 & $V$ &  19.17 $\pm$ 0.08 &  $300 \times 4 $ & 5 \\ \\

56712.62 & 2014 Feb 24 & $R_c$& 20.48 $\pm$ 0.12  & $140 \times 5 $ & 6 \\ 
56726.51 & 2014 Mar 10 & $R_c$& 19.48 $\pm$ 0.19  &  $60 \times 27$ & 2 \\
56726.75 & 2014 Mar 10 & $R_c$& 19.73 $\pm$ 0.06  & $140 \times 7$  & 6 \\
56726.93 & 2014 Mar 10 & $R_c$& 20.01 $\pm$ 0.10  &  $300 \times 4 $ & 5 \\ 
56731.53 & 2014 Mar 15 & $R_c$ &  19.58 $\pm$ 0.46  &  $60 \times 26$ & 2 \\
56732.53 & 2014 Mar 16 & $R_c$ &  19.48 $\pm$ 0.39  &  $60 \times 54$ & 2 \\ \\

56726.51 & 2014 Mar 10 & $I_c$ & 18.89 $\pm$ 0.48   &  $60 \times 27$ & 2 \\
56726.77 & 2014 Mar 10 & $I_c$ & 19.41 $\pm$ 0.10   &  $140 \times 7$   & 6 \\ 
56731.53 & 2014 Mar 15 & $I_c$ & 18.50 $\pm$ 0.20   &  $60 \times 26$ & 2 \\
56732.53 & 2014 Mar 16 & $I_c$ & 18.91 $\pm$ 0.19  &  $60 \times 54$ & 2 \\ \\

56712.60 & 2014 Feb 24 & $J$  & 18.74 $\pm$ 0.60 &   $120 \times 5$  &  6  \\
56726.75 & 2014 Mar 10 & $J$  & 18.10 $\pm$ 0.11 &   $120 \times 7$  &  6  \\
56732.68 & 2014 Mar 16 & $J$  & 17.43 $\pm$ 0.07 &   $120 \times 9$  &  6  \\
56733.61 & 2014 Mar 17 & $J$  & 17.72 $\pm$ 0.11 &   $120 \times 12$ &  6 \\ \\

56726.77 & 2014 Mar 10 & $K_s$ & 16.01 $\pm$ 0.07 &  $120 \times 7$  & 6  \\
56732.70 & 2014 Mar 16 & $K_s$ & 15.84 $\pm$ 0.08 &  $120 \times 9$  & 6  \\

\hline \hline
         & UT         & Filter &  Magnitude     & Source  & Offset$^{e}$ \\ \hline
         & 1966      & $B$   & 20.09            &  USNO-B1.0 & 0.6''  \\
         & 1966      & $B$   & 20.58            & USNO-B1.0  & 0.6''  \\
         & 1994      & $Bj$  & 22.06  $\pm$ 0.65 & GSC-II    & 0.3'' \\
         & 1994      & $B$   & 20.73  $\pm$ 0.51 & GSC-II    & 0.3'' \\
         & 2003 Mar 24 & $u$ &  22.76 $\pm$ 0.40 & SDSS      & \\
         &            & $g$ &  22.16 $\pm$ 0.11 &  SDSS    & \\
         &            & $r$ &  22.17 $\pm$ 0.13 &  SDSS    & \\
         &            & $i$ &  21.91 $\pm$ 0.16 &  SDSS    & \\
         &            & $z$ &  20.94 $\pm$ 0.25 &  SDSS    & \\ 
         &  2010      & $W1$ &  17.41 $\pm$ 0.14  &   WISE & 0.36'' \\ 
         &            & $W2$ &  16.93 $\pm$ 0.32  &   WISE & 0.36'' \\ 
         &            & $W3$ &  $>12.78^f$          &  WISE & 0.36'' \\ 
         &            & $W4$ &  $>8.63^f$           &  WISE & 0.36'' \\ \hline \hline 
         &    UT      & Frequency &  Flux    &  Source$^g$ & Offset$^{e}$ \\
         &           &    (GHz) &  (mJy)    &               &   \\ \hline
         & 1976-1978   & 0.151          &  91 $\pm$ 13  &   7C    & 0.11'  \\
         & 1991-1996   & 0.325          &  47            &   WENSS & 0.46' \\
         & 1993-1997   & 1.4            &  21.3 $\pm$ 0.8 &   NVSS & 0.04' \\
         & 1995-1998   & 1.4            &  15.76 $\pm$ 0.13 &  FIRST & 0.003' \\
         & 1986-1987   & 4.85           &  31 $\pm$ 4  &   GB6  & 0.047' \\
         & 1995 Aug 14 & 8.4            &  18.8        &   CLASS & 0.004' \\
\enddata
\label{tab:data}
\end{deluxetable*}

\setcounter{table}{0}

\begin{deluxetable*}{cccccc} 
\tablewidth{0pt}
\tablecaption{continued}
\tablehead{ & & & & & }
\startdata
\multicolumn{2}{c}{Line}  &  \multicolumn{2}{c}{Flux} &  \multicolumn{2}{c}{Luminosity} \\
\multicolumn{2}{c}{}      &  \multicolumn{2}{c}{(${\rm 10^{-17} \ erg \ s^{-1}\ cm^{-2}}$)} &  \multicolumn{2}{c}{(${\rm 10^{41} \ erg \ s^{-1}}$)}  \\ \hline

\multicolumn{2}{c}{\ion{Mg}{ii} 2798}    & \multicolumn{2}{c}{48.9  $\pm$ 2.8}  & \multicolumn{2}{c}{15.6  $\pm$ 0.9}   \\
\multicolumn{2}{c}{[\ion{Ne}{V}] 3425}   & \multicolumn{2}{c}{7.7  $\pm$ 0.7}  & \multicolumn{2}{c}{2.5  $\pm$ 0.2}    \\
\multicolumn{2}{c}{[\ion{O}{ii}] 3727}   & \multicolumn{2}{c}{17.0  $\pm$ 1.6}  & \multicolumn{2}{c}{5.4  $\pm$ 0.5}     \\
\multicolumn{2}{c}{[\ion{Ne}{iii}] 3869} & \multicolumn{2}{c}{10.2  $\pm$ 0.3}  & \multicolumn{2}{c}{3.3  $\pm$ 0.1}    \\
\multicolumn{2}{c}{[\ion{Ne}{iii}] 3968} & \multicolumn{2}{c}{5.2 $\pm$  0.7}  & \multicolumn{2}{c}{1.7   $\pm$ 0.2}    \\
\multicolumn{2}{c}{H$\gamma$ (narrow)}   & \multicolumn{2}{c}{5.8  $\pm$ 0.6}  & \multicolumn{2}{c}{1.9  $\pm$ 0.2}    \\
\multicolumn{2}{c}{H$\beta$ (narrow)}    & \multicolumn{2}{c}{10.3  $\pm$ 1.8}  & \multicolumn{2}{c}{3.3   $\pm$ 0.6}     \\
\multicolumn{2}{c}{H$\beta$ (broad)}     & \multicolumn{2}{c}{16.2  $\pm$ 2.7}  & \multicolumn{2}{c}{5.2  $\pm$ 0.9}    \\
\multicolumn{2}{c}{[\ion{O}{iii}] 4959}  & \multicolumn{2}{c}{30.9  $\pm$ 1.0}  & \multicolumn{2}{c}{9.9  $\pm$ 0.3}    \\
\multicolumn{2}{c}{[\ion{O}{iii}] 5007}  & \multicolumn{2}{c}{93.5  $\pm$ 1.4}  & \multicolumn{2}{c}{29.9  $\pm$ 0.4}    \\
\enddata
\tablecomments{
$^a$ Average time for stacked data. 
$^b$ AB magnitudes for SDSS $ugriz$ filters
and Vega magnitudes for the other filters. 
$^c$ 1. Kiso/KWFC, 2. MITSuME, 3. OAO/KOOLS, 4. Subaru/FOCAS, 
5. Kottamia Observatory, 6. Kanata/HONIR
$^d$ $5 \sigma$ upper limit.
$^e$ Offset from the source position in the SDSS images.
$^f$ 95 \% confidence upper limit in AllWISE Source Catalog.
$^g$ The survey/catalog abbreviations and references: 
Seventh Cambridge Survey \citep[7C,][]{hales07},
Westerbork Northern Sky Survey \citep[WENSS,][]{rengelink97},
Faint Images of the Radio Sky at Twenty-Centimeters
\citep[FIRST,][]{white97}, 
NRAO VLA Sky Survey \citep[NVSS,][]{condon98}, 
Green Bank 6-cm Radio Source Catalog \citep[GB6,][]{gregory96},
The Cosmic Lens All-Sky Survey \citep[CLASS,][]{myers03}.
}
\label{tab:data2}
\end{deluxetable*}

\end{document}